\documentclass[runningheads]{svmult}

\usepackage{makeidx}   
\usepackage{graphicx}  
\usepackage{subeqnar}  
\usepackage{multicol}  
\usepackage{physprbb}  
\makeindex             

\newcommand{\greeksym}[1]{{\usefont{U}{psy}{m}{n}#1}}
\newcommand{\umu}{\mbox{\greeksym{m}}}
\newcommand{\degree}{\mbox{$^{\circ}$}}
\newcommand{\arcmin}{\mbox{$^{\prime}$}}
\newcommand{\arcsec}{\mbox{$^{\prime\prime}$}}

\begin{document}
\title*{R=100,000 Spectroscopy of Photodissociation Regions: 
H$_2$ Rotational Lines in the Orion Bar}
\toctitle{R=100,000 Spectroscopy of Photodissociation Regions: 
H$_2$ Rotational Lines in the Orion Bar}

%
%
\titlerunning{H$_2$ Rotational Lines in the Orion Bar}
%
\author{Katelyn N. Allers\inst{1,3}
\and Daniel T. Jaffe\inst{1,3}
\and John H. Lacy\inst{1,3}
\and Matthew J. Richter\inst{2,3}}
\authorrunning{Allers et al.}
%
%
\institute{Department of Astronomy, University of Texas at Austin, Austin, TX 78712-0259
\and Department of Physics, University of California, Davis, 1 Shields Avenue, Davis, CA 95616
\and Visiting Astronomer at the Infrared Telescope Facility, which is operated by the University of Hawaii under Cooperative Agreement no. NCC 5-538 with the National Aeronautics and Space Administration, Office of Space Science, Planetary Astronomy Program.}

\maketitle              


\section{Introduction}

Photodissociation regions (PDRs) form on the surfaces of 
molecular clouds whenever these clouds are struck by far-ultraviolet 
radiation from hot young stars.  These regions are 
characterized by the transition from hot, 
ionized gas to cold, molecular gas as the far-ultraviolet field is attenuated 
farther from the ionization front.  
The 
temperature profile of the PDR varies depending on the density and strength 
of the FUV field.  Derivation of this profile must take into account the 
local heating and cooling, the chemical equilibrium, and radiative coupling 
to other layers within the structure.
The Orion Bar is
a dense molecular structure at the southeast boundary of
the Orion Nebula.  
Early theoretical models of the Orion Bar by Tielens and Hollenbach 
\cite{tielens} (for a density of 2.3 $\times$ 10$^4$ and UV field strength 
(G$_0$) of 10$^5$) predict 
temperatures of $\sim$1000 K at A$_{\mathrm{V}}$=0-2, dropping to 
less than 100 K by A$_{\mathrm{V}}$=4.  

Ground state rotational lines of H$_2$ are good temperature 
probes of moderately hot 
(200-1000 K) gas.  The low A-values of these lines result in low critical 
densities while ensuring that the lines are optically 
thin.  ISO observations of H$_2$ rotational lines in PDRs reveal large 
quantities of warm gas that are difficult to explain via current 
models\cite{draine}, but 
the spatial resolution of ISO does not resolve the 
temperature structure of the warm gas.  We present and discuss high spatial 
resolution observations of H$_2$ rotational line emission from the Orion Bar.

\section{Observations and Data Reduction}
We mapped the H$_2$ $v$ = 0-0 S(1) and S(2) lines at 17.03 \umu m 
and 12.28 \umu m toward the Orion 
Bar in 2002 December. We made the observations using the Texas Echelon 
Cross Echelle Spectrograph (TEXES, \cite{lacy}) on the 3m NASA Infrared 
Telescope Facility (IRTF).  
The slit width was 2.0\arcsec\ for the 0-0 S(1) line and 1.4\arcsec\ for the 
0-0 S(2) line.  The spectral resolution of our 
data is 5.5 km s$^{-1}$ at 17.03 \umu m 
and 4.7 km s$^{-1}$ at 12.28 \umu m.
On source integration times per position were 160 and 
80 seconds for the 0-0 S(1) and S(2) lines respectively.  
We oriented the slit parallel to the Orion Bar ionization 
front, at a position angle of 45\degree. 
We mapped by stepping the telescope from northwest to southeast 
in 0.7\arcsec\ (for S(2)) and 1\arcsec\ (for S(1)) steps to 
create 40\arcsec\ long scans. 
We reduced the raw images of cross-dispersed spectra using the standard TEXES 
pipeline reduction program \cite{lacy}. 
We then smoothed our data to a spatial resolution of 2\arcsec. 
The 0,0 position for the maps is at 
R.A. = 5$^h$ 35$^m$ 19.7$^s$, Dec. = -5\degree\ 25\arcmin\ 
28.3\arcsec\ (J2000.0) and the mapped region runs from 13\arcsec\ 
northwest to 27\arcsec\ southeast of this position.

\section{Results}

\subsection{Morphology}

\begin{figure}[t]
\begin{center}
\includegraphics[width=\textwidth]{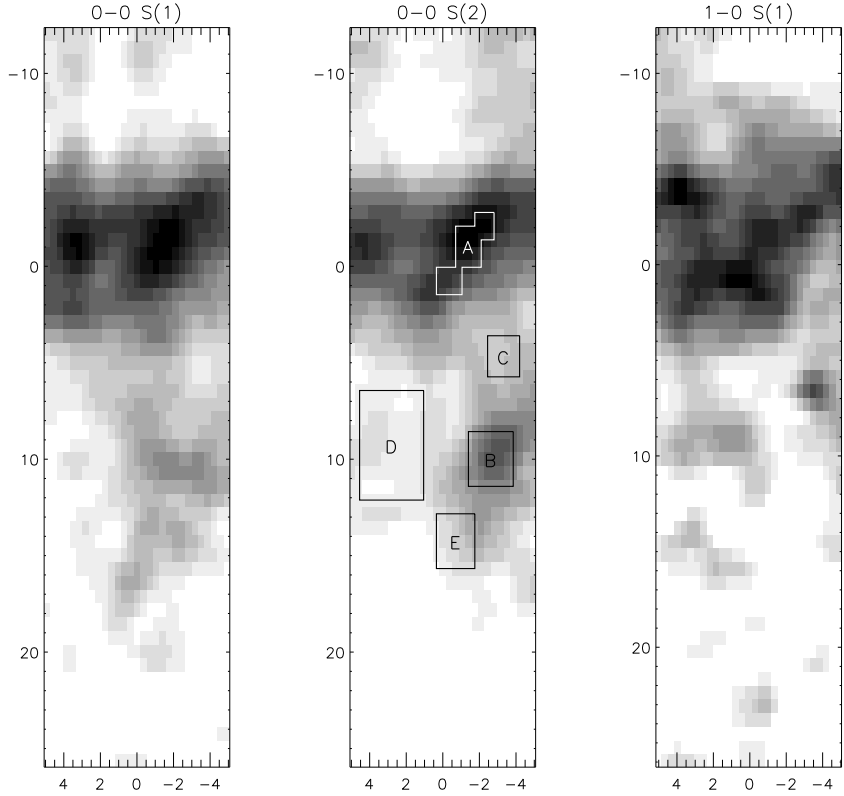}
\end{center}
\caption[]{Maps of integrated intensities for the S(1) and S(2) pure 
rotational lines taken with TEXES along with the $v$ = 
1-0 S(1) line \cite{vanderwerf}.  Labelled axis are in arcseconds. 
The scans were taken perpendicular to the ionization front at a position angle 
of 45\degree.  The 0,0 position in the maps 
corresponds to R.A. = 5$^h$ 35$^m$ 19.7$^s$, Dec. = -5\degree\ 25\arcmin\ 
28.3\arcsec\ (J2000.0).  The wedge of the greyscale goes from the 1 sigma 
noise to the maximum value of the maps, or 
0.33 to 9.6 $\times$ 10$^{-4}$ ergs cm$^{-2}$ s$^{-1}$ sr$^{-1}$ for 0-0 S(1),  
0.22 to 8.2 $\times$ 10$^{-4}$ ergs cm$^{-2}$ s$^{-1}$ sr$^{-1}$ for 0-0 S(2), and 
0.25 to 4.4 $\times$ 10$^{-4}$ ergs cm$^{-2}$ s$^{-1}$ sr$^{-1}$ for 1-0 S(1).  
The middle panel labels the areas averaged together to increase 
signal-to-noise for further analysis.}
\label{maps}
\end{figure}
Figure \ref{maps} shows the distribution of $v$=0-0 S(1) and S(2) intensity,
along with distribution of intensity in the 2.12 
\umu m $v$ = 1-0 S(1) line \cite{vanderwerf}, 
resampled onto the same
grid at the same spatial resolution. 
The dominant feature in the maps is the bright horizontal (northeast-southwest)
ridge centered at y$\simeq$-2\arcsec.  
This feature has a sharp northwestern edge, rising
at most positions from a low or undetectable level to half of its peak 
intensity in $\leq$2\arcsec\ (0.004 pc).  
The total 
thickness of the bright ridge is $\simeq$10\arcsec\ (0.02 pc).

There is remarkable agreement in the intensity distributions, not only
between the 0-0 S(1) and 0-0 S(2) lines but also between these lines and the
$v$=1-0 S(1) line.  The maps 
agree to within the 
uncertainties about the location and width of the bright ridge, and
about the presence and extent of lower-level extended emission. 
To the left of its center, the bright ridge contains a clumpy 
``V'' structure whose overall morphology echoes in all three lines.

\subsection{Temperatures and Column Densities}
Table 1 lists the temperatures and column densities for the areas labeled in 
Figure \ref{maps}.  Derived excitation temperatures range from roughly 
400 to 700 K, with most areas 
at around 600 K.  Over the 20\arcsec\ range in depth where we 
are able to measure temperature, we do not see any systematic trend in 
temperature with depth into the molecular cloud (to the southeast and farther 
from $\theta^{1}$C).

\begin{table}
\begin{center}
\renewcommand{\arraystretch}{1.4}
\caption{Physical Conditions}
\begin{tabular}{lllllll}
\hline\noalign{\smallskip}
& \multicolumn{3}{c}{Observed Intensities} & & & \\
Area & 0-0 S(1) & 0-0 S(2) & 1-0 S(1) & T$_{\mathrm{ex}}$$^{\mathrm a}$ & T$_{\mathrm{ex}}$$^{\mathrm b}$ & N(H$_2$)$^{\mathrm c}$\\ 
& \multicolumn{3}{c}{$10^{-4}$ erg cm$^{-2}$ s$^{-1}$} & K & K & 10$^{20}$ cm$^{-2}$\\
\noalign{\smallskip}
\hline
A & 7.7 & 6.6 & 3.6  & 476 & 630 & 7.7  \\
B & 3.8 & 4.3 & 0.49 & 591 & 554 & 3.1  \\
C & 1.8 & 2.1 & 1.3  & 616 & 661 & 1.4  \\
D & 1.1 & 1.3 & 0.94 & 668 & 673 & 0.81 \\
E & 2.6 & 1.8 & $<$0.2& 413 & $<$525& 3.1  \\
\noalign{\smallskip}
\hline
\end{tabular}
\end{center}
$^{\mathrm a}$ Excitation temperature determined from I[0-0 S(1)]/I[0-0 S(2)] assuming optically thin emission with o/p=3. \\
$^{\mathrm b}$ Excitation temperature determined from I[1-0 S(1)]/I[0-0 S(1)] assuming optically thin emission with o/p=3. \\
$^{\mathrm c}$ Column density of warm H$_2$ for thermalized gas at the temperature determined from I[0-0 S(1)]/I[0-0 S(2)]. \\ 
\label{tab1}
\end{table}

At most positions the temperature derived from the 0-0 S(1)/0-0 S(2)
line intensity ratio agrees to better than 50 K with the temperature 
derived from the 1-0 S(1)/0-0 S(1) line intensity ratio despite a factor of 4 
difference in E$_{upper}$ between the 0-0 S(2) and 1-0 S(1) transitions.  
Looking at the entire dataset, we see the most variability 
between the two derived temperatures close to the ionization front. 
Because of the small variation in temperature over the mapped region 
of the PDR, the 
0-0 S(2) line intensity map roughly traces the column density of warm H$_2$. 

\subsection{Line widths and velocities}
With the spectral resolution available with TEXES, we are able to resolve the 
ground vibrational state H$_2$ lines.
Physical linewidths (after 
deconvolving the instrument profile) are 3-6 km s$^{-1}$, in agreement 
with linewidths expected for optically thin, thermalized gas at 600 K. 
Line widths for the 
two transitions agree to within 1-2 km s$^{-1}$, with neither line 
being systematically wider than the other.  
We measure V$_{LSR}$ for the H$_2$ lines to be 10-11 km s$^{-1}$.
Our observed line widths and 
velocities are in agreement with published values for both 
molecular lines (NH$_3$ \cite{batrla}, CS, and isotopes of 
CO \cite{hogerheijde}) and carbon recombination lines \cite{wyrowski}.

\section{Discussion}
PDR models predict that H$_2$ emission arises at a depth in the PDR with 
steep temperature gradients \cite{tielens,draine}.  
We do not see evidence for such spatial gradients even though we have 
resolved the PDR.  The agreement of temperatures derived 
from the 0-0 S(1)/0-0 S(2) and 1-0 S(1)/0-0 S(1) line intensity ratio 
indicates that the emitting gas is isothermal.  
We re-interpret the 400-700 K temperatures found for galactic PDRs in the 
large ISO beam \cite{wright} as emerging from isothermal gas, and not the 
beam-average of a temperature gradient.  
Future PDR modeling efforts must be able 
to explain the ubiquity of 600 K H$_2$ in PDRs.

%

\end{document}